\documentclass[aps,onecolumn]{revtex4}%
\usepackage{amsfonts}
\usepackage{amsmath}
\usepackage{amssymb}
\usepackage{graphicx}%
\newtheorem{theorem}{Theorem}

\newenvironment{proof}[1][Proof]{\noindent\textbf{#1.} }{\ \rule{0.5em}{0.5em}}
\begin{document}
\title{The von Neumann entropy of networks}
\author{Filippo Passerini}
\affiliation{Perimeter Institute for Theoretical Physics, Waterloo, Ontario N2L 2Y5, Canada
and Department of Physics and Astronomy, University of Waterloo, Ontario N2L
3G1, Canada}
\author{Simone Severini}
\affiliation{Institute for Quantum Computing and Department of Combinatorics \&
Optimization, University of Waterloo, Waterloo N2L 3G1, ON Canada}

\begin{abstract}
We normalize the combinatorial Laplacian of a graph by the degree sum, look at
its eigenvalues as a probability distribution and then study its Shannon
entropy. Equivalently, we represent a graph with a quantum mechanical state
and study its von Neumann entropy. At the graph-theoretic level, this quantity
may be interpreted as a measure of regularity; it tends to be larger in
relation to the number of connected components, long paths and nontrivial
symmetries. When the set of vertices is asymptotically large, we prove that
regular graphs and the complete graph have equal entropy, and specifically it
turns out to be maximum. On the other hand, when the number of edges is fixed,
graphs with large cliques appear to minimize the entropy.

\end{abstract}
\maketitle

\section{Introduction}

The \emph{quantum entropy} (or, equivalently, \emph{von Neumann entropy}) was
defined by von Neumann around 1927 for proving the irreversibility of quantum
measurement processes \cite{v}. Precisely, the quantum entropy is an extension
of the Gibbs entropy to the quantum realm and it may be viewed as the average
information the experimenter obtains in the repeated observations of many
copies of an identically prepared mixed state. It has a fundamental role for
studying correlated systems and for defining entanglement measures \cite{nc,
o}. In the present work we elaborate on the notion of quantum entropy applied
to networks. Since the quantum entropy is defined for quantum states, the
first required ingredient is therefore a method to map graphs/networks into
states (while the converse is not necessary in our purpose). The literature
comprises different ways to associate graphs to certain states or dynamics.
Notably, graph-states and spin networks, just to mention two major ones:
graph-states are certain quantum error correcting codes, important for
characterizing the computational resources in measurement based quantum
computation \cite{br, h}; spin networks are arrangements of interacting
quantum mechanical particles, nowadays of great significance for the
development of nanotechnologies \cite{bl, bos, ki}. We take a straightforward
approach, and take into analysis an entropic quantity for graphs on the basis
of a faithful mapping between discrete Laplacians and quantum states, firstly
introduced by Braunstein \emph{et al. }\cite{b} (see also \cite{hms}). In
synthesis, we see the spectrum of an appropriately normalized Laplacian as a
distribution and we compute its Shannon entropy \cite{ct} (which measures the
amount of uncertainty of a random variable, or the amount of information
obtained when its value is revealed). Such a quantity finds a natural place
among those \emph{global} spectral parameters of graphs (\emph{i.e.},
involving the entire spectrum and not just a specific eigenvalue) studied in
connection to natural and social networks. For example, the Estrada index, a
measure of centrality \cite{fre}, also used to quantify the degree of folding
of long-chain molecules \cite{es1, es2, laz}; or the \emph{graph energy}, that
in H\"{u}ckel theory corresponds to the sum of the energies of all the
electrons in a molecule \cite{cds, gu}. (See the book chapter \cite{bb}, for a
general review on complexity measures for graphs.) We give evidence that the
quantum entropy is a measure of regularity for graphs, \emph{i.e.}, regular
graphs have in general higher entropy when the number of edges is fixed.
Moreover, entropy seems to depend on the number of connected components, long
paths, and nontrivial symmetries. Chosen the number of edges, entropy is
smaller for graphs with large cliques and short paths, \emph{i.e.}, graphs in
which the vertices form an highly connected cluster. The remainder of the
paper is organized as follows. In Section \ref{two} we introduce the required
definitions and focus on some basic properties. By adding edges one by one to
the empty graph, we try to construct graphs with minimum and maximum entropy.
In Section \ref{egs} we explore the influence of the graph structure on the
entropy. We consider different classes of graphs: regular graphs, random
graphs, and the star as an extremal case of scale-free graph (\emph{i.e.},
graphs for which the degree distribution follows a power law). The asymptotic
behavior for large number of vertices shows that regular graphs tend to have
maximum entropy. We study numerically how the entropy increases when adding
edges with different prescriptions. Once fixed the number of edges, the
entropy is minimized by graphs with large cliques. Section \ref{conclusion}
contains remarks and open problems.

\section{First properties\label{two}}

The state of a quantum mechanical system with a Hilbert space of finite
dimension $n$ is described by a \emph{density matrix}. Each density matrix
$\rho$ is a positive semidefinite matrix with Tr$(\rho)=1$. As we have already
mentioned in the introduction, there are many ways to associate graphs to
specific density matrices or Hamiltonian evolution (\emph{e.g.}, graph states,
bosonic systems, \emph{etc.}). Here we consider a matrix representation based
on the combinatorial Laplacian. Let $G=(V,E)$ be a simple undirected graph
with set of vertices $V(G)=\{1,2,...,n\}$ and set of edges $E(G)\subseteq
V(G)\times V(G)-\{\{v,v\}:v\in V(G)\}$. The \emph{adjacency matrix} of $G$ is
denoted by $A(G)$ and defined by $[A(G)]_{u,v}=1$ if $\{u,v\}\in E(G)$ and
$[A(G)]_{u,v}=0$, otherwise. The \emph{degree} of a vertex $v\in V(G)$,
denoted by $d(v)$, is the number of edges adjacent to $v$. A graph $G$ is
$d$-\emph{regular} if $d(v)=d$ for all $v\in V(G)$. Let $d_{G}$ be the
\emph{degree-sum} of the graph, \emph{i.e.} $d_{G}=\sum_{v\in V(G)}d(v)$. The
\emph{average degree} of $G$ is defined by $\bar{d}_{G}:=\tilde{n}^{-1}%
\sum_{v\in V(G)}d(v)$, where $\tilde{n}$ is the number of \emph{non-isolated
vertices}, that is vertices $v$ such that $\{u,v\}\in E(G)$ for some $u\in
V(G)$. The \emph{degree} \emph{matrix} of $G$ is an $n\times n$ matrix,
denoted by $\Delta(G)$, having $uv$-th entry defined as follows: $\left[
\Delta(G)\right]  _{u,v}=d(v)$ if $u=v$ and $\left[  \Delta(G)\right]
_{u,v}=0$, otherwise. The \emph{combinatorial Laplacian} \emph{matrix }of a
graph $G$ (for short, \emph{Laplacian}) is the matrix $L(G)=\Delta(G)-A(G)$.
The matrix $L(G)$ is a major tool for enumerating spanning trees (via the
Matrix-Tree Theorem) and has numerous applications (see Kirchhoff \cite{kir},
Biggs \cite{big}, and Grone \emph{et al.} \cite{g, g1}). As a consequence of
the Ger\v{s}gorin disc theorem, $L(G)$ is positive semidefinite. By these
definitions, the Laplacian of a graph $G$ scaled by the degree-sum of $G$ is a
density matrix: $\rho_{G}:=\frac{L(G)}{d_{G}}=\frac{L(G)}{\text{Tr}%
(\Delta(G))}$. It is then clear that $\rho_{G}=\frac{L(G)}{\tilde{n}\bar
{d}_{G}}$. The entropy of a density matrix $\rho$ is defined as $S(\rho
)=-$Tr$(\rho\log_{2}\rho)$. Now, given the notion of Laplacian, we say that
$S(\rho_{G})$ is\emph{ the quantum entropy }(or, for short, \emph{entropy})
\emph{of }$G$. Let $\nu_{1}\geq\nu_{2}\geq\cdots\geq\nu_{n}=0$ and
$\lambda_{1}\geq\lambda_{2}\geq\cdots\geq\lambda_{n}=0$ be the eigenvalues of
$L(G)$ and $\rho_{G}$, respectively. These are related by a scaling factor,
\emph{i.e.} $\lambda_{i}=\frac{\nu_{i}}{d_{G}}=\frac{\nu_{i}}{\tilde{n}\bar
{d}_{G}}$, for $i=1,...,n$. The entropy of $\rho_{G}$ can be also written as
$S(G)=-\sum_{i=1}^{n}\lambda_{i}\log_{2}\lambda_{i}$, where $0\log_{2}0=0$, by
convention. (See \cite{mo} for a survey on Laplacian spectra.) Since its rows
sum up to $0$, then $0$ is the smallest eigenvalue of $\rho_{G}$. The number
of connected components of $G$ is equal to the multiplicity of $0$ as an
eigenvalue. The largest Laplacian eigenvalue is bounded by the number of
non-isolated vertices, \emph{i.e.}, $\nu_{1}\leq\tilde{n}$ (see Duval \emph{et
al.} \cite{dr}, Proposition 6.2); thus it follows immediately that
$0\leq\lambda_{i}\leq\frac{1}{\bar{d}_{G}}$, for $i=1,...,n$. It is important
to remark that since $0\leq\lambda_{i}\leq\frac{1}{\bar{d}_{G}}\leq1$, for
$i=1,...,n$, $S(\rho_{G})$ equals the Shannon entropy of the probability
distribution $\{\lambda_{i}\}_{i=1}^{n}$. If a general density matrix $\rho$
has an eigenvalue $1$ then the other must be $0$ and $\rho=\rho^{2}$. In such
a case, the density matrix is said to be \emph{pure}; otherwise, \emph{mixed}.
For later convenience, we define the quantity $R(G):=\frac{1}{n}\sum_{i=1}%
^{n}\frac{\nu_{i}}{\bar{d}_{G}}\log_{2}\frac{\nu_{i}}{\bar{d}_{G}}$. The
\emph{disjoint union }of graphs $G$ and $H$ is the graph $G^{\prime}=G\uplus
H$, whose connected components are $G$ and $H$. We denote by $K_{n}$ the
complete graph on $n$ vertices. Let $\mathcal{G}_{n}$ be the set of all graphs
on $n$ vertices. The next fact was proved by Braunstein \emph{et al.} \cite{b}:

\begin{theorem}
Let $G$ be a graph on $n\geq2$ vertices. Then $\min_{\mathcal{G}_{n}}S(G)=0$
if and only if $G=K_{2}\biguplus_{j}K_{1}^{(j)}$ and $\max_{\mathcal{G}_{n}%
}S(G)=\log_{2}(n-1)$ if and only if $G=K_{n}$. When $n=2$, then $\min
_{\mathcal{G}_{2}}S(G)=\max_{\mathcal{G}_{n}}S(G)=0$ and $G=K_{2}$.
\end{theorem}

For general density matrices, $S(\rho)=0$, if $\rho$ is a pure state;
$S(\rho)=-\log_{2}\frac{1}{n}=\log_{2}n$ if $\rho=\frac{1}{n}I_{n}$,
\emph{i.e.}, a completely random state. The analogue in $\mathcal{G}_{n}$ is
$K_{n}$ given that the spectrum or $\rho_{K_{n}}$ is $\{\frac{1}{n-1}%
^{[n-1]},0^{[1]}\}$. The next result bounds the variation of the entropy under
edge addition. Let $G^{\prime}=G+\{x,y\}$, where $V(G)=V(G^{\prime})$ and
$E(G^{\prime})=E(G)\cup\{u,v\}$. An alternative proof could be given by
invoking eigenvalues interlacing \cite{cds}.

\begin{theorem}
\label{conc}For graphs $G$ and $G^{\prime}=G+\{x,y\}$, we have $S(\rho
_{G^{\prime}})\geq\frac{d_{G^{\prime}}-2}{d_{G^{\prime}}}S(\rho_{G})$.
\end{theorem}

\begin{proof}
Chosen a labeling of $V(G)$, for $G\in\mathcal{G}_{n}$ we can write
$A(G)=\sum_{\{u,v\}\in E(G)}A(u,v)$, where $A(u,v)$ is the adjacency matrix of
a graph $G(u,v):=\{u,v\}\biguplus_{i=1}^{n-2}K_{1}$. We can then define an
$n\times n$ diagonal matrix $\Delta(u,v)$ such that $[\Delta(u,v)]_{u,u}%
=[\Delta(u,v)]_{v,v}=1$ and $[\Delta(u,v)]_{u,v}=0$ if $u\neq v$. It follows
that $\Delta(G)=\sum_{\{u,v\}\in E(G)}\Delta(u,v)$. Then $\rho_{G}=\frac
{1}{d_{G}}\sum_{\{u,v\}\in E(G)}(\Delta(u,v)-A(u,v))$. Let $\{|1\rangle
,|2\rangle,...,|n\rangle\}$ be the standard basis of $\mathbb{C}^{n}$. By
definition, $|i\rangle\equiv(0_{1},0_{2}...,0_{i-1},1_{i},0_{i+1}%
,...,0_{n})^{T}$. We associate the pure state $|\{u,v\}\rangle=\frac{1}%
{\sqrt{2}}(|u\rangle-|v\rangle)$ to the edge $\{u,v\}$. Let $P(u,v)$ be the
projector associated to $|\{u,v\}\rangle$: $P(u,v)=\frac{1}{2}(I_{2}%
-\sigma_{x})$. Then $\rho_{G}=\frac{2}{d_{G}}\sum_{\{u,v\}\in E(G)}P(u,v)$ and
$\rho_{G^{\prime}}=\frac{d_{G}}{d_{G^{\prime}}}\rho_{G}+\frac{2}{d_{G^{\prime
}}}P(x,y)$. It is well-known that the entropy $S$ is concave (see Ohya and
Petz \cite{o}): $S\left(  \sum_{i=1}^{l}\alpha_{i}\rho_{i}\right)  \geq
\sum_{i=1}^{l}\alpha_{i}S(\rho_{i})$, where $\rho_{i}$ are density matrices
and $\alpha_{i}\in\mathbb{R}^{+}$. Hence $S(\rho_{G^{\prime}})\geq\frac{d_{G}%
}{d_{G^{\prime}}}S(\rho_{G})+\frac{2}{d_{G^{\prime}}}S(\{x,y\})$. However,
since $S(\{x,y\})=0$, the claim is true.
\end{proof}

\begin{figure}
[h]
\begin{center}
\includegraphics[
height=2.1482in,
width=3.4281in
]%
{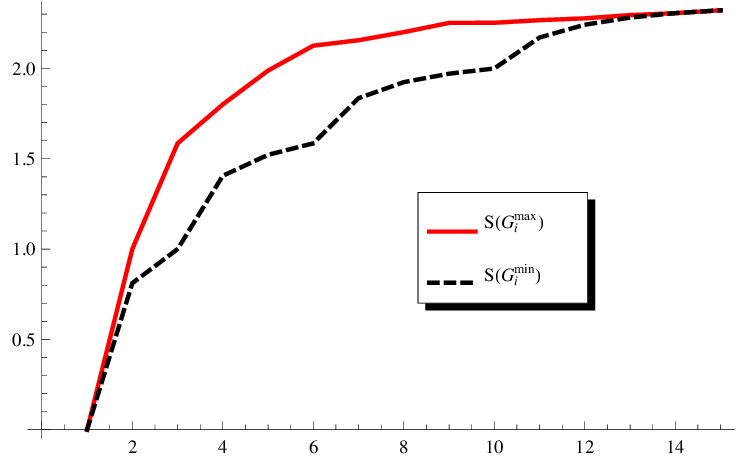}%
\caption{Plots of $S(G_{i}^{\max})$ and $S(G_{i}^{\min})$ (resp. solid and
dashed line) as functions of the number of edges $i=1,2,...,15$. }%
\label{min}%
\end{center}
\end{figure}

\begin{figure}[hh]
\begin{center}
\includegraphics[
width=3.397in
]{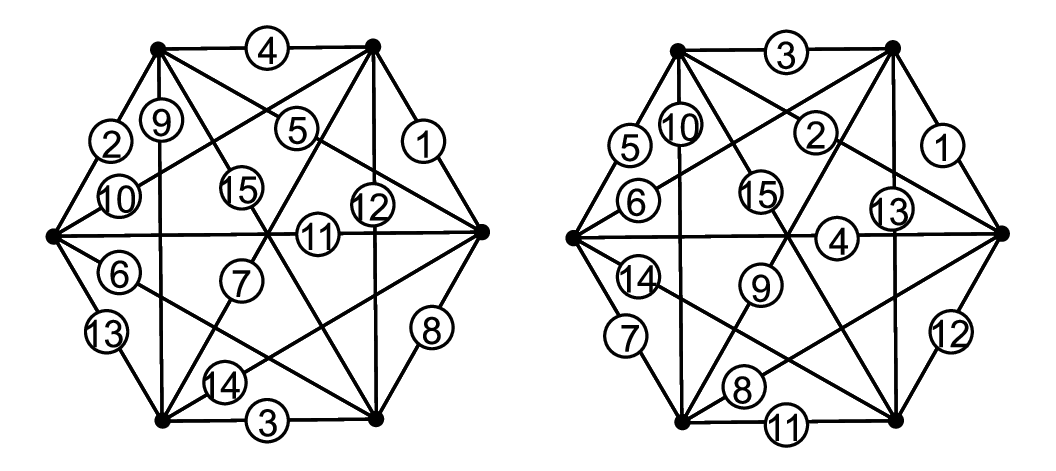}
\end{center}
\caption{This figure shows two complete graphs $K_{6}$ with labeled edges. For
the graph on the left hand side, the edge labeled by $i$ is added at time $i$
in order to construct $G_{i}^{\max}$. The graph on the right hand side is the
analogue drawing for $G_{i}^{\min}$. }%
\label{ksix}%
\end{figure}

Starting from $K_{2}\biguplus_{j}K_{1}^{(j)}$ (the graph with zero entropy) we
can think of a discrete-time process in which we add edges so that the entropy
is extremal (resp. maximum or minimum) at every step. Let us denote by
$G_{i}^{\max}$ and $G_{i}^{\min}$, $i\geq1$, the graphs with maximum and
minimum entropy at the $i$-th step, respectively. Figure \ref{min} contains
$S(G_{i}^{\max})$ and $S(G_{i}^{\min})$ (resp. solid and dashed line) as
functions of the number of edges $i=1,2,...,15$, for graphs in $\mathcal{G}%
_{6}$. The initial graph is $G_{1}^{\max}=G_{1}^{\min}=K_{2}\biguplus
_{j=3}^{6}K_{1}^{(j)}$; the final one is $G_{15}^{\max}=G_{15}^{\min}=K_{6}$.
Each edge labeled by $j\leq i$ in the graph $K_{6}$ on the left (resp. right)
hand side of Figure \ref{ksix} is also an edge of $G_{i}^{\max}$ (resp.
$G_{i}^{\min}$). This illustrates the steps for constructing every
$G_{i}^{\max}$ and $G_{i}^{\min}$. It turns out that the vertices of
$G_{i}^{\max}$ tend to have \textquotedblleft almost equal\textquotedblright%
\ or equal degree. In fact $G_{i}^{\max}$ is a $i/3$-regular graph, for
$i=3,6,9,12$. On the other hand, $G_{l(l-1)/2}^{\min}=K_{l}\biguplus
_{j=1}^{6-l}K_{1}^{(j)}$, if $l=3,4,5$. The meaning is without ambiguity:
entropy is minimized by those graphs with \emph{locally }added edges,
\emph{i.e.} edges increasing the number of complete subgraphs (also called
\emph{cliques}). Even if we consider graphs with only six vertices, it is
already evident that long paths, nontrivial symmetries and connected
components give rise to a larger increase of the entropy. This property is
confirmed by further numerical analysis in the next section.%


\section{Entropy and graph structure\label{egs}}

Let $\mathcal{G}_{n,d}$ be the set of all $d$-regular graphs. For
$G\in\mathcal{G}_{n,d}$, we have $\Delta(G)=dI_{n}$, and hence $\lambda
_{i}=\frac{d-\mu_{i}}{\text{Tr}\left(  \Delta(G)\right)  }=\frac{d-\mu_{i}%
}{dn}$, for $i=1,2,...,n$, where $\mu_{i}$ denotes the $i$-th eigenvalue of
$A(G)$.

\begin{theorem}
\label{th1}Let $G$ be a graph on $n$ nonisolated vertices. If $\lim
_{n\rightarrow\infty}\frac{R(G)}{\log_{2}n}=0$ then $\lim_{n\rightarrow\infty
}\frac{S(G)}{S(K_{n})}=1$. In particular, if $G\in\mathcal{G}_{n,d}$ then
$\lim_{n\rightarrow\infty}\frac{S(G)}{S(K_{n})}=1$.
\end{theorem}

\begin{proof}
When $G\in\mathcal{G}_{n}$, $\rho_{G}=\frac{L(G)}{n\overline{d}_{G}}$, where
$\bar{d}_{G}=\frac{1}{n}\sum_{v\in V(G)}d(v)$. Since $\lambda_{i}=\frac
{\nu_{i}}{n\bar{d}_{G}}$, we have $S(G)=-\frac{1}{n}\sum_{i=1}^{n}\frac
{\nu_{i}}{\bar{d}_{G}}\log_{2}\frac{\nu_{i}}{\bar{d}_{G}}+\frac{1}{n}%
\sum_{i=1}^{n}\frac{\nu_{i}}{\bar{d}_{G}}\log_{2}n$. Given that Tr$(\rho
_{G})=\frac{\nu_{i}}{n\bar{d}_{G}}=1$, by taking $R(G):=\frac{1}{n}\sum
_{i=1}^{n}\frac{\nu_{i}}{\bar{d}_{G}}\log_{2}\frac{\nu_{i}}{\bar{d}_{G}}$, the
quantum entropy of $G$ is given by $S(G)=-R(G)+\log_{2}n$. Since
$S(K_{n})=\log_{2}(n-1)$, we have $S(G)=-R(G)+\frac{S(K_{n})\log_{2}n}%
{\log_{2}(n-1)}$. From this expression, we see immediately that if
$\lim_{n\rightarrow\infty}\frac{R(G)}{\log_{2}n}=0$ then $\lim_{n\rightarrow
\infty}\frac{S(G)}{S(K_{n})}=1$. Now, let us consider $G\in\mathcal{G}_{n,d}$.
Since $d(v)=d$ for every $v\in V(G)$, it follows that $\bar{d}=d$ and $\nu
_{i}=d-\mu_{i}$. Given that for a $d$-regular graph $-d\leq\mu_{i}\leq d$, we
have $0\leq\nu_{i}\leq2d$, for every $i=1,...,n$. The quantity $R(G)$ is now
given by $R(G)=\frac{1}{n}\sum_{i=1}^{n}x_{i}\log_{2}x_{i}$, where
$x_{i}=\frac{\nu_{i}}{d}$, and $0\leq x_{i}\leq2$. The function $x_{i}\log
_{2}x_{i}$ assumes finite values in the range $[0,2]$. Thus $R(G)$ is also
finite. In particular, since $R(G)$ is an average, it remains finite even if
considering an arbitrary large number of vertices. This implies that the
entropy for a $d$-regular graph tends to the entropy of $K_{n}$ in the limit
$n\rightarrow\infty$.
\end{proof}

It may useful to remark two points: \emph{(1) }The simplest regular graph is
the \emph{perfect matching }$M_{n}:=\biguplus_{j=1}^{n/2}K_{2}^{(j)}$. The
density matrix of $M_{n}$ is then $\rho_{M_{n}}=\frac{1}{n}\bigoplus
\nolimits_{n/2\text{ times}}\left(
\begin{array}
[c]{rr}%
1 & -1\\
-1 & 1
\end{array}
\right)  $ and $S(M_{n})=-\frac{n}{2}\left(  \frac{2}{n}\log_{2}\frac{2}%
{n}\right)  =\log_{2}\frac{n}{2}$, because $\lambda_{1}^{[\frac{n}{2}]}%
=\frac{n}{2}$ and $\lambda_{2}^{[\frac{n}{2}]}=0$. Thus, $S(M_{n}%
)=S(K_{n/2+1})$. For $M_{4}$ we have $S(M_{4})=S(K_{3})=1$. More generally,
$S\left(  M_{2^{k}}\right)  =k-1$. \emph{(2) }The entropy of $G\in
\mathcal{G}_{n\rightarrow\infty}$ tends to the entropy of $K_{n}$ if all the
quantities $\frac{\nu_{i}}{\bar{d}_{G}}$ remain finite, \emph{i.e.},
$\lim_{n\rightarrow\infty}\frac{R(G)}{\log_{2}n}=0$.

The \emph{complete bipartite graph} $K_{p,q}$ has $V(K_{p,q})=A\cup B$, where
$\left\vert A\right\vert =p$ and $\left\vert B\right\vert =q$, and each vertex
in $A$ is adjacent to every vertex in $B$. The graph $K_{1,n-1}$ on $n$
vertices is said to be a \emph{star}.

\begin{theorem}
Let $G\in\mathcal{G}_{n}$ with $v$ such that $\{v,u\}\in E(G)$ for every $u$,
and let $\lim_{n\rightarrow\infty}\bar{d}_{G}=d_{\infty}<\infty$. Then
$\lim_{n\rightarrow\infty}\frac{S(G)}{S(K_{n})}\leq1-{\frac{1}{d_{\infty}}}$.
In particular, the star $K_{1,n-1}$ saturates the bound, since $d_{\infty}=2$,
and $\lim_{n\rightarrow\infty}S(K_{1,n-1})/S(K_{n})=\frac{1}{2}$.
\end{theorem}

\begin{proof}
Let $G$ be as in the statement. So, $d_{1}=n-1$. For a graph with at least one
edge, Grone \emph{et al.} (see \cite{g1}, Corollary 2) proved that $\nu
_{1}\geq d_{1}+1$; for a generic graph on $\tilde{n}=n$ vertices, we know that
$\nu_{1}\leq n$ (see Duval \emph{et al.} \cite{dr}, Proposition 6.2). By these
two results, $\nu_{1}=n$. Thus, we have $R(G)=\frac{1}{n}\sum_{i=1}^{n}%
\ \frac{\nu_{i}}{\bar{d}_{G}}\log_{2}\frac{\nu_{i}}{\bar{d}_{G}}=\frac{1}%
{\bar{d}_{G}}\log_{2}\frac{n}{\bar{d}_{G}}+\frac{1}{n}\sum_{i=2}^{n}%
\ \frac{\nu_{i}}{\bar{d}_{G}}\log_{2}\frac{\nu_{i}}{\bar{d}_{G}}$ and
$\lim_{n\rightarrow\infty}\frac{R(G)}{S(K_{n})}\geq{\frac{1}{d_{\infty}}}$.
Because $S(G)=-R(G)+\log_{2}n$, we have $\lim_{n\rightarrow\infty}\frac
{S(G)}{S(K_{n})}\leq1-{\frac{1}{d_{\infty}}}$. Now, the eigenvalues of
$\rho_{K_{1,n-1}}$ are $\lambda_{1}^{[1]}=\frac{n}{2n-2}$, $\lambda
_{2}^{[n-2]}=\frac{1}{2n-2}$ and $\lambda_{3}^{[1]}=0$. Thus, the entropy is
given by $S(K_{1,n-1})=-\frac{n}{2n-2}\log_{2}\frac{n}{2n-2}+\frac{n-2}%
{2n-2}\log_{2}(2n-2)$ and in the limit $n\rightarrow\infty$ we have the second
part of the statement. Since $\bar{d}=\frac{2n-2}{n}$, it results $d_{\infty
}=2$ and the bound is saturated.
\end{proof}

\begin{figure}[h]
\begin{center}
\includegraphics[
height=2.0609in,
width=3.397in
]{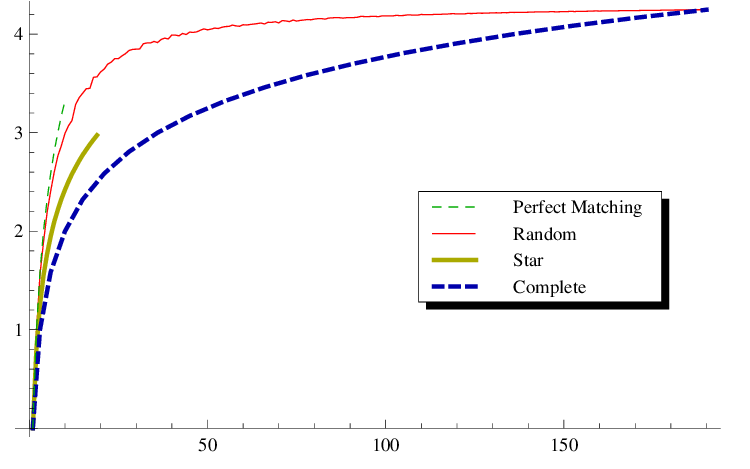}
\end{center}
\caption{Plots of the entropy of four different kind of graphs as a function
of the number of edges $e=1,2,\ldots,190$. The different plots represent
different ways of adding edges to a graph with $n=20$ vertices. The value of
$S(R_{n,e})$ has been avaraged over 15 different random graphs, for each value
of $e$.}%
\label{plot}%
\end{figure}

Similarly to what we have done in the previous section, we observe how the
entropy of a graph $G\in\mathcal{G}_{n}$ increases as a function of
$\left\vert E(G)\right\vert =e$. Starting from $K_{2}\biguplus_{j=1}%
^{n-2}K_{1}^{(j)}$, we consider four different ways of adding edges: \emph{(i)
}Random graphs with exactly $e$ edges. These are constructed by chosing $e$
pairs of vertices at random from the total number of pairs; \emph{(ii) }The
graph $M_{2e}\biguplus_{j=1}^{n-2e}K_{1}^{(j)}$; \emph{(iii) }The graph
$K_{1,(e+1)-1}\biguplus_{j=1}^{n-e-1}K_{1}^{(j)}$; \emph{(iv) }The graph
$K_{m}\biguplus_{j=1}^{n-m}K_{1}^{(j)}$, where $m=\left[  \frac{1+\sqrt{1+8e}%
}{2}\right]  $. Recall that adding isolated vertices to a graph does not
change its entropy. Figure \ref{plot} shows the case $n=20$. It is evident
that the entropy is larger for graphs with an high number of connected
components. In this sense, $M_{n}$ has relatively high entropy. The smallest
entropy is obtained for complete graphs.

\section{Conclusions\label{conclusion}}

Next is a list of remarks and open problems:

\noindent\emph{Normalized Laplacian. }We have considered the combinatorial
laplacian $L(G)$. There is a related matrix called \emph{normalized Laplacian}
and defined by $\mathcal{L}(G)=\Delta^{-1/2}L(G)\Delta^{-1/2}$ (by convention
$[\Delta^{-1}]_{v,v}=0$ if $d(v)=0$). It results that $[\mathcal{L}%
(G)]_{u,v}=1$ if $u=v$ and $d(v)\neq0$, $[\mathcal{L}(G)]_{u,v}=-1/\sqrt
{d(u)d(v)}$ if $\{u,v\}\in E(G)$, and $[\mathcal{L}(G)]_{u,v}=0$, otherwise
(see \cite{fc, te}). If a graph has no isolated vertex then Tr$(\mathcal{L}%
(G))=n$. Therefore, we can define the density matrix $\widehat{\rho}%
_{G}:=\frac{\mathcal{L}(G)}{n}$. The entropy of $\widehat{\rho}_{G}$ is then
$S(\widehat{\rho}_{G})=-$Tr$(\frac{\mathcal{L}(G)}{n}\log_{2}\frac
{\mathcal{L}(G)}{n})=-W+\log_{2}n=-\frac{1}{n}$Tr$\left(  \mathcal{L}%
(G)\log_{2}\mathcal{L}(G)\right)  +\log_{2}n$. Since the eigenvalues of
$\mathcal{L}(G)$ are in $[0,2]$ \cite{fc}, when limit $n\rightarrow\infty$,
the quantity $W$ remains finite. We may then conclude that when the number of
vertices goes to infinity, the entropy $S(\widehat{\rho}_{G})$ tends to
$S(\rho_{K_{n}})$. This fact provides a motivation for dealing with $L(G)$
instead of $\mathcal{L}(G)$.

\noindent\emph{Algebraic connectivity. }Let $a(G)=\nu_{n-1}$ be the
\emph{algebraic connectivity }of $G$ \cite{fie}. It is nonzero only if $G$ is
connected. The value of $a(G)$ quantifies the connectivity of $G$. Is there a
relation between $a(G)$ and $S(\rho_{G})$? Consider $K_{n}$ and the
$n$-\emph{cycle} $C_{n}$, that is the connected $2$-regular graph on $n$
vertices. For these, $a(K_{n})=n$ and $a(C_{n})=2(1-\cos\frac{2\pi}{n})$. By
Theorem \ref{th1}, $\lim_{n\rightarrow\infty}S(C_{n})=S(K_{n})$. However the
algebraic connectivity of the two graphs behave differently in this limit:
$\lim_{n\rightarrow\infty}a(K_{n})=\infty$ and $\lim_{n\rightarrow\infty
}a(C_{n})=0$.

\noindent\emph{Eigenvalue gap. }Let $b(G)=\mu_{1}-\mu_{2}$ be the
\emph{eigenvalue gap }of $G$. This parameter determines the mixing time of a
simple random walk on $G$ (see Lov\'{a}sz \cite{lo}). If $G\in\mathcal{G}%
_{n,d}$ then $a(G)=b(G)$. Hence $\lim_{n\rightarrow\infty}b(K_{n})=\infty$ and
$\lim_{n\rightarrow\infty}b(C_{n})=0$. We can therefore state that $b(G)$ and
$S(\rho_{G})$ describe different properties of $G$ at least on the basis of
this basic observation.

\noindent\emph{A combinatorial definition. }It is unclear whether $S(G)$ is
related to combinatorially defined entropic quantities. For example, the
K\"{o}rner entropy defined in \cite{ko} (see also Simonyi \cite{si} for a
survey) or the entropies defined by Riis \cite{ri} and Bianconi \cite{ginb}.
Intuitively, any relation should be weak, because the quantum entropy depends
on the eigenvalues. For this reason it describes some global statistical
behaviour, with only partial control over combinatorial properties.

\noindent\emph{Beyond cospectrality. }Graphs with the same eigenvalues have
equal entropy. We have seen that also perfect matchings and complete graphs
plus a specific number of isolated vertices have equal entropy, but are
clearly noncospectral (see Section \ref{egs}). Determine families of graphs
with the same entropy remains an open problem.

\noindent\emph{Relative entropy. }The quantum relative entropy is a measure of
distinguishability between two states (see the review \cite{vv}). Given two
graphs $G$ and $H$, the \emph{quantum relative entropy} may be defined as
$S(G||H):=-$Tr$(\rho_{G}\log_{2}\rho_{H})-S(\rho_{G})$. What kind of relations
between the two graphs are emphasized by the relative entropy? To what extent
can this be used as a measure of distinguishability for graphs?

We conclude with two open problems: does the star $K_{1,n-1}$ have smallest
entropy among all connected graphs on $n$ vertices? Is the entropy strictly
monotonically increasing under edge addition?

\bigskip

\noindent\emph{Acknowledgments. }The authors would like to thank Michele
Arzano, Alioscia Hamma, Dan Lynch, Yasser Omar, Federico Piazza and Samuel
Vazquez, for helpful discussion. FP was supported by the Perimeter Institute
for Theoretical Physics. Research at Perimeter is supported by the Government
of Canada through Industry Canada and by the Province of Ontario through the
Ministry of Research \& Innovation. SS was supported by the Institute for
Quantum Computing. Research at the Institute for Quantum Computing is
supported by DTOARO, ORDCF, CFI, CIFAR, and MITACS.

\end{document}